\documentclass[11pt]{cernrep}
\usepackage{cite}
\usepackage{amsmath}
\usepackage{amssymb}
\usepackage{graphicx}
\usepackage{here}

\begin{document}
\bibliographystyle{unsrt}
\def\D0{D\O~}
\renewcommand{\thefootnote}{\fnsymbol{footnote}}

\begin{flushright}
hep-ph/0401171
\end{flushright}

\noindent{\Large\bf Combined Effect of QCD Resummation and QED Radiative
Correction to $W$ boson Mass Measurement at the LHC\footnote{submitted to the proceedings of the Workshop on
Physics at TeV Colliders, Les Houches, France, 
26 May -- 6 June 2003}}\\[8mm]
\textit{Qing-Hong Cao and C.-P. Yuan}\\
{Department of Physics and Astronomy, Michigan State University, USA}

\begin{abstract}
A precise determination of the $W$ boson mass at the CERN LHC
requires a theoretical calculation in which the effect of the initial-state
multiple soft-gluon emission and the final-state photonic correction
are simultaneously included . Here, we present such a calculation
and discuss its prediction on a few most relevant distributions of
the decay leptons.
\end{abstract}

\section{INTRODUCTION}
As a fundamental parameter of the Standard Model (SM), the mass of the
$W$-boson ($M_{W}$) is of particular importance. Aside from being
an important test of the SM itself, a precision measurement of $M_{W}$,
together with an improved measurement of top quark mass ($M_{t}$),
provides severe indirect bounds on the mass of Higgs boson
($M_{H}$). With a precision of 15 MeV for $M_{W}$~\cite{Haywood:1999qg} and 2 GeV for
$M_{t}$ at the LHC~\cite{unknown:1999fr},  $M_{H}$ in the SM
can be predicted with an uncertainty of about 30\%~\cite{Haywood:1999qg}.
Comparison of these indirect constraints on $M_{H}$ with the results
from direct Higgs boson searches, at the 
LEP2, the Tevatron and the CERN
Large Hadron Collider (LHC), will be an important test of the SM. In order
to have a precision measurement of $M_{W}$, the theoretical uncertainties,
dominantly coming from the transverse momentum of the $W$-boson ($P_{T}^{W}$),
the uncertainty in parton distribution function (PDF) and the
electroweak (EW) radiative
corrections to the $W$ boson decay, must be controlled
to a better accuracy~\cite{Baur:2000bi,Haywood:1999qg}.

At the LHC, most of $W$ boson is produced in the small transverse momentum region.
When $P_{T}^{W}$ is much smaller than $M_{W}$,
every soft-gluon emission will induce a large
logarithmic contribution to the $P_{T}^{W}$ distribution so that the
order-by-order perturbative calculation
in the theory of Quantum chromodynamics (QCD)
cannot accurately describe the
$P_{T}^{W}$ spectrum
and the contribution from multiple soft-gluon emission,
which contributes to all orders in the expansion of the strong coupling
constant $\alpha_s$, needs to be summed to all orders.
It has been shown that by applying the renormalization group
analysis, the multiple soft-gluon radiation effects can be resummed
to all orders to predict the $P_{T}^{W}$ distribution that agrees
with experimental data~\cite{Balazs:1995nz, Balazs:1997xd}. RESBOS,
a Monte Carlo (MC) program ~\cite{Balazs:1997xd}
resumming the initial-state soft-gluon radiations
of the hadronically produced lepton pairs through EW vector boson
production and decay at hadron colliders
$p\bar{p}/pp\rightarrow V(\rightarrow\ell_{1}\bar{\ell_{2}})X$,
has been used by the CDF and \D0 Collaborations at the Tevatron to
compare with their data in order to determine $M_W$.
However, RESBOS does not include any
higher order EW corrections to describe the vector boson decay. The EW
radiative correction, in particular the final-state QED
correction, is crucial for precision measurement of $W$ boson mass
at the Tevatron, because photon emission from the final-state charged
lepton can significantly modify the lepton momentum which is used in
the determination of $M_{W}$. In the CDF Run Ib $W$ mass measurement,
the mass shifts due to radiative effects were estimated to be $-65\pm20$
MeV and $-168\pm10$ MeV for the electron and muon channels, respectively~\cite{Affolder:2000bp}.
The full next-to-leading order (NLO) $O(\alpha)$ EW
corrections have been calculated~\cite{Dittmaier:2001ay, Baur:1998kt}
and resulted in WGRAD~\cite{Baur:1998kt}, a MC program
for calculating $O(\alpha)$ EW radiative corrections to the process
$p\bar{p}\rightarrow\nu_{\ell}\ell(\gamma)$. However, WGRAD does not include
the dominant correction originated from the initial-state multiple soft-gluon
emission. To incorporate both the initial-state QCD and and final-state
QED corrections into a parton level MC program is urgently required
to reduce the theoretical uncertainties in interpreting the experimental
data at the Tevatron. It was shown in Refs. \cite{Dittmaier:2001ay, Baur:1998kt} that at the NLO,
the EW radiative correction in $p\bar{p}\rightarrow\ell\nu_{l}(\gamma)$ is
dominated by the final-state QED (FQED) correction. Hence, in this
paper we present a consistent calculation which includes both the
initial-state multiple soft-gluon QCD resummation and the final-state
NLO QED corrections, and develop an upgraded version of
the RESBOS program,
called RESBOS-A~\footnote{A Fortran code that implemets the theoretical calculation presented in this work.},
to simulate the signal events. Here, we only present
the phenomenological impacts on a few experimental observalbes, the transverse mass of $W$ boson ($M_T^W$) and
the transverse momentum of charged lepton ($p_T^{\ell}$), that are most sensitive to the
measurement of $M_W$. We focus our attention on
the electron lepton only, though our analysis procedure also applies
to the $\mu$ lepton. The detailed formula, the SM inputting parameters and the kinematics cuts
are given in Ref.~\cite{resbosa}.

\section{PRECISION MEASUREMENT OF W MASS}
\begin{figure}
\begin{center}
\includegraphics[width=10cm]{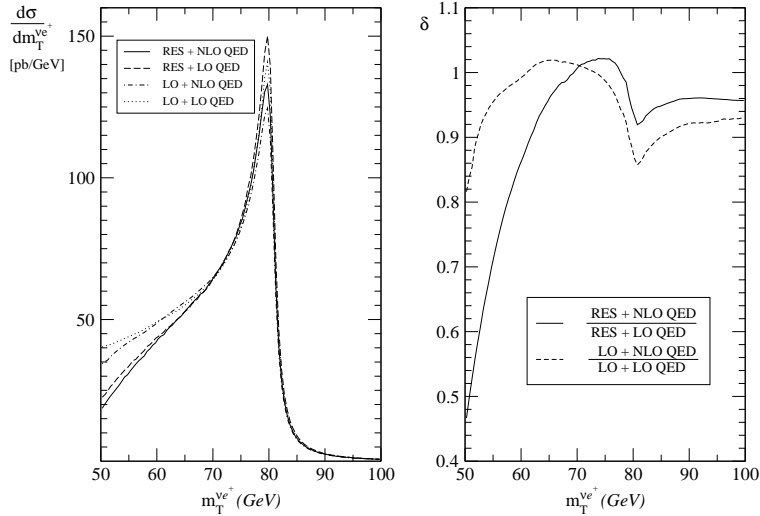}
\caption{Transverse mass distribution of $W^+$ boson \label{fig:tm}}
\end{center}
\end{figure}

In Fig.~\ref{fig:tm}, we show various theory predictions on
the $M_{T}^{W}$ distribution.
The legend of the figure is defined as follows:
\begin{itemize}
\item LO : including only the Born level initial-state contribution,
\item RES : including the initial-state multiple soft-gluon corrections
via QCD resummation,
\item LO QED : including only the Born level final-state contribution,
\item NLO QED : including the final-state NLO QED corrections.
\end{itemize}
For example, the solid curve (labelled as RES+NLO QED) in
Fig.~\ref{fig:tm}(a) is the prediction from our combined calculation, 
given by Eqs.~(1) and (2) of
Ref~\cite{resbosa}.

As shown in Fig.~\ref{fig:tm}(a), compared to the lowest order cross
section (dotted curve), the initial state QCD resummation effects (dashed
curve) increase the cross section at the peak of the $M_{T}^{W}$ distribution
by about $6\%$, and the final state NLO QED corrections (dot-dashed
curve) decrease it by about $-12\%$, while the combined
contributions (solid curve)
of the QCD resummation and FQED corrections reduce it by $6\%$.
In addition to the change in magnitude,
the line-shape of the $M_{T}^{W}$ distribution
is significantly modified by the effects of QCD resummation
and FQED corrections. 
To illustrate this point, we plot the ratio of the (RES+NLO QED)
differential cross sections to the LO ones
as the solid curve in Fig.~\ref{fig:tm}(b).
The dashed curve is for the ratio of (LO+NLO QED) to LO.
As shown in the figure,
the QCD resummation effect dominates the shape of
$M_{T}^{W}$ distribution for
$65\,{\rm GeV\leq M_{W}\leq95\,{\rm GeV}}$, while
the FQED correction reaches its maximal effect around the Jacobian
peak ($M_{T}^{W}\simeq M_{W}$).
Hence, both corrections must be included to accurately predict
the distribution of $M_{T}^{W}$ around the Jacobian region to determine
$M_W$. We note that after including the effect due to the finite
resolution of the detector (for identifying an isolated electron or
muon), the size of the FQED correction is largely 
reduced~\cite{Dittmaier:2001ay, Baur:1998kt}.
\begin{figure}
\begin{center}
\includegraphics[width=10cm]{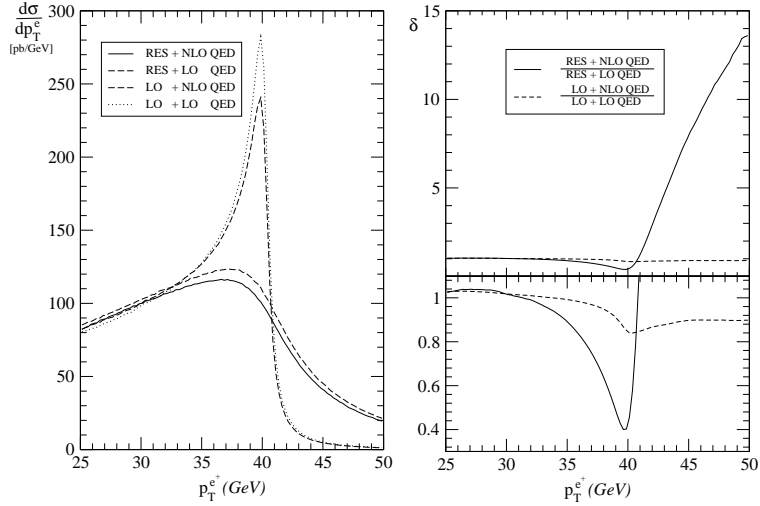}
\caption{Transverse momentum distributions of $e^+$\label{fig:pte}}
\end{center}
\end{figure}

Although the $M_{T}^{W}$ distribution is an optimal observable
for determining $M_{W}$ at the LHC with a low luminosity, it requires
 an accurate measurement of the missing transverse momentum
direction which becomes more difficult to control with a high luminosity 
option (when multiple scattering becomes important).
On the other hand, the transverse momentum
of the decay charged lepton ($p_{T}^{e}$) is less sensitive to
the detector resolution, so that it can be used to measure $M_W$
and provide important cross-check on the result derived from
the $M_{T}^{W}$ distribution,
for they have different systematic uncertainties. 
Another important feature of this observable is that
$p_{T}^{e}$ distribution is more sensitive to the transverse momentum
of $W$ boson. Hence, the QCD soft-gluon resummation effects,
the major source of $p_{T}^{W}$, must be included to reduce
the theoretical uncertainty of this method.
In Fig.~\ref{fig:pte}(a), we show the $p_{T}^{e}$
distributions predicted by various theory calculations, 
and in Fig.~\ref{fig:pte}(b), the ratios
of the higher order to lowest order cross sections as a function
of $p_{T}^{e}$. The lowest order distribution
(dotted curve) shows a clear and sharp Jacobian peak at 
$p_{T}^{e}\simeq M_{W}/2$,
and the distribution with the 
NLO final-state QED correction (dot-dashed curve) also
exhibits the similar Jacobian peak with the peak magnitude
reduced by about $15\%$. But the clear and sharp Jacobian peak of
the lowest order and NLO FQED distributions 
(in which $p_{T}^{W}=0$)
are strongly smeared by
the finite transverse momentum of the $W$ boson induced by multiple
soft-gluon radiation, as clearly demonstrated by the 
QCD resummation distribution
(dashed curve) and the combined contributions of the QCD resummation and
FQED corrections (solid curve).
Similar to the $M_{T}^{W}$ distribution, the QCD resummation effect
dominates the whole $p_{T}^{e}$ range, while the FQED correction
reaches it maximum around the Jacobian peak (half of $M_W$).
The combined contribution of the QCD resummation
and FQED corrections reaches the order of $45\%$ near 
the Jacobian peak. Hence, these lead us to conclude that the
QCD resummation effects are crucial in the measurement of $M_{W}$
from fitting the Jacobian kinematical edge of the $p_{T}^{e}$
distribution. 

In order to study the impact of the presented calculation to the determination
of the $W$ boson mass, the effect due to the finite resolution of
the detector should be included, which will be presented elsewhere.

\section{ACKNOWLEDGEMENTS}
We thank P.~Nadolsky and J.~W.~Qiu for helpful discussions.  This work
was supported in part by NSF under grand No. PHY-0244919 and PHY-0100677.
\bibliography{biblifile}

\begin{thebibliography}{1}

\bibitem{Haywood:1999qg}
S.~Haywood et~al.
\newblock Electroweak physics.
\newblock 1999.

\bibitem{unknown:1999fr}
Atlas detector and physics performance. technical design report. vol. 2.
\newblock CERN-LHCC-99-15.

\bibitem{Baur:2000bi}
U.~Baur, R.~K. Ellis, and D.~Zeppenfeld.
\newblock Qcd and weak boson physics in run ii. proceedings, batavia, usa,
  march 4-6, june 3-4, november 4-6, 1999.
\newblock Prepared for Physics at Run II: QCD and Weak Boson Physics Workshop:
  Final General Meeting, Batavia, Illinois, 4-6 Nov 1999.

\bibitem{Balazs:1995nz}
Csaba Balazs, Jian-wei Qiu, and C.~P. Yuan.
\newblock Effects of qcd resummation on distributions of leptons from the decay
  of electroweak vector bosons.
\newblock {\em Phys. Lett.}, B355:548--554, 1995.

\bibitem{Balazs:1997xd}
C.~Balazs and C.~P. Yuan.
\newblock Soft gluon effects on lepton pairs at hadron colliders.
\newblock {\em Phys. Rev.}, D56:5558--5583, 1997.

\bibitem{Affolder:2000bp}
T.~Affolder et~al.
\newblock Measurement of the w boson mass with the collider detector at
  fermilab.
\newblock {\em Phys. Rev.}, D64:052001, 2001.

\bibitem{Dittmaier:2001ay}
Stefan Dittmaier and Michael Kramer.
\newblock Electroweak radiative corrections to w-boson production at hadron
  colliders.
\newblock {\em Phys. Rev.}, D65:073007, 2002.

\bibitem{Baur:1998kt}
U.~Baur, S.~Keller, and D.~Wackeroth.
\newblock Electroweak radiative corrections to w boson production in hadronic
  collisions.
\newblock {\em Phys. Rev.}, D59:013002, 1999.

\bibitem{resbosa}
Qing-Hong Cao and C.-P. Yuan.
\newblock MSUHEP-040106, hep-ph/0401026.

\end{thebibliography}
\end{document}